\documentstyle[12pt]{article}

\begin{document}

		\begin{titlepage}
\setcounter{page}{1}

\date{}
\title{\small\centerline{\hfill September 1993 }
\bigskip\bigskip
{\Large\bf Unstable particle mixing and CP violation in weak decays}
         \bigskip}

\author{ Jiang Liu and Gino Segr\`e \\
\normalsize
\em Department of Physics\\
\normalsize
\em University of Pennsylvania\\
\normalsize
\em Philadelphia, PA 19104\\
\normalsize
UPR-0582T}
\maketitle
\vfill

\begin{abstract}\noindent\normalsize
We discuss unstable particle mixing
in CP-violating weak decays.
It is shown that
for a completely degenerate system
unstable particle mixing does not introduce a
CP-violating partial rate difference, and
that when the mixings are small only the
off-diagonal mixings are relevant.  Also,
in the absence of mixing, unstable particle wave function
renormalization does not introduce any additional effect.
An illustrative example is given to heavy scalar decays
with arbitrary mixing.
\end{abstract}

\vfill

\thispagestyle{empty}
\end{titlepage}


\section{ Introduction}

The smallness of $K_L-K_S$ mass difference allows us to have
an access to
rare processes such as CP violation.
Up to now the only established experimental evidence of CP violation
comes from the mixing of
the unstable particles $K^0$ and $\bar{K}^0$ \cite{Winstein} .

Earlier studies of unstable particle mixing followed two
physically equivalent paths.  One is due to Weisskopf and Wigner \cite{WW},
in which one introduces an effective  complex mass matrix.
The  evolution of the system is
determined by the standard time-dependent
Hamiltonian formalism \cite{LOY}.
The other
is due to Sachs \cite{Sachs1}, in which  one studies the
dynamics of the complex pole of the kaon field propagator.
The Hamiltonian method is expressed directly
in terms of the measured quantities and is therefore more
transparent from a phenomenological viewpoint.
On the other hand,
the propagator method  arises naturally in the context
of quantum field theory, and hence is more easily
adapted to fundamental gauge theories of weak interactions.
Both approaches are  phenomenological, having difficulties
handling ultraviolet
divergences arising from higher order corrections.
In spite of these fundamental difficulties, the phenomenological
formalisms have been very successful.  They  provide
the standard descriptions for the study of unstable particle mixing.

The advent of renormalizable gauge theory provides a connection between
the parameters of a phenomenological formalism and the
parameters of a given fundamental theory.
In this paper we would like to study these connections for
unstable particle mixing in some detail,  focusing
on CP-violating processes.  We will
 adopt an approach that combines
the two methods mentioned above.

Instead of introducing
a complete renormalization prescription, our immediate goal
is more modest. In the next section we discuss some general
properties of $S$-matrix elements in the presence  of unstable
particle mixing.  The results of this analysis turn
out to be very useful for simplifying Feynman diagram calculations.

In section 3 we study the relationship  between the unstable particle
mixing and antiparticle mixing.  For simplicity, we
only focus on scalars.    A simple formula
valid for small mixings is derived for
CP-violating partial rate differences.

The formalisms developed in section 2 and section 3
 are  applied to a simple
example of baryogenesis by heavy scalar decay.
The results are shown to agree
with the published results obtained directly from Feynman
diagram calculations.  This part
is presented in section 4, followed
by a discussion in section 5 of large mixing and renormalization.
Our conclusion is presented in section 6.
We give two  appendices to present some technical details:
one discusses the renormalization of unstable particle
mixing and
the other shows how to diagonalize an arbitrary
$n\times n$ complex matrix.
\bigskip

\section{ General formalism}

Consider the weak decay
of a  set of unstable particles  $\phi_a$  produced
at $t=0$, where the index  $a=1,2,...$
labels different flavor of $\phi$.   Suppose the
lowest order  amplitude of $\phi_a$ decaying into a
final state $|F_{f}\rangle$  is given by
\begin{eqnarray}
T_{fa}=\langle F_{f}|H_{\rm weak}|\phi_a\rangle.\label{Telement}
\end{eqnarray}
By CPT invariance, to first order of $H_{\rm weak}$
 a replacement of $F_{f}$ and $\phi_a$
by their antiparticles $\bar{F}_{f}$ and $\bar{\phi}_a$
corresponds
to change $T_{fa}$ to $T_{fa}^*$.
By the superposition principle,
the weak amplitudes at a later time $t$ are
\begin{eqnarray}
T_{fa}(t)&=&\sum_{b,c}T_{fb}V_{bc}^{-1}V_{ca}e^{-i\omega_c t},
\label{amplitude}\\
\bar{T}_{fa}(t)& =&\sum_{b,c}T_{fb}^*\bar{V}_{bc}^{-1}\bar{V}_{ca}
e^{-i\omega_c t},\label{antiamplitude}
\end{eqnarray}
where $V$ and $\bar{V}$ are the mixing matrices
\begin{eqnarray}
|\phi_a\rangle = V_{ca}|\phi_c'\rangle,\label{mixingmatrix}\\
|\bar{\phi}_a\rangle=\bar{V}_{ca}|\bar{\phi}_c'\rangle,\label{antimixing}
\end{eqnarray}
and $|\phi'_c\rangle,~
|\bar{\phi}'_c\rangle$ are the eigenstates of
an effective Hamiltonian, which is correct up to second
order in $H_{\rm weak}$.  $|\phi'_c\rangle$ and $|\bar{\phi}'_c\rangle$
will be referred to
as the eigenstate of propagation hereafter. The eigenvalue of
 $|\phi'_c\rangle$ and $|\bar{\phi}'_c\rangle$ is
$\omega_c=m_c-i\gamma_c/2$ in the rest frame $\phi'_c$, where $m_c$
and $\gamma_c$ may be interpreted, respectively, as the mass and width of
$\phi_c'$ and $\bar{\phi}_c'$.

If $|F_{f}\rangle\ (|\bar{F}_{f}\rangle)$ belongs to
the unstable particle set,
then $T_{fb}$ and $\bar{T}_{fb}$ are zero unless $f=b$, and
Eqs. (\ref{amplitude}) and (\ref{antiamplitude})
are useful for the study of  time distribution of CP asymmetry in oscillation
\cite{Cabi,Kuo}.  Except for situations in which the
particles are stable or the mass and decay matrices commute,
$V$ and $\bar V$ are in general  not unitary.
If both CPT  and CP are conserved, $V=\bar V$  and
Eqs.  (\ref{amplitude}) and (\ref{antiamplitude})
are  the natural  generalizations of the known formalism  \cite{FHF} describing
the so called `mix-and-decay' phenomena.
The formalism of \cite{FHF}
has been employed in the study of unstable neutrino oscillation
\cite{BC,APP}.  For practical purpose the mixing matrix
has so far been approximated as unitary.  Such an approximation
is not always justifiable.

Our main interest is in the unstable particle mixing effect in
the time-integrated rate difference
\begin{eqnarray}
\Delta_{fa}=\Gamma(\phi_a\to F_{f})-\Gamma(\bar{\phi}_a\to \bar{F}_{f}).
\label{ratedifference}
\end{eqnarray}
It should be pointed out that,
in addition to the mixing, the rate difference may also
depends on final-state interactions,
and it is not always
possible to separate them if these two effects are comparable.
A discussion on particle mixing is
always warranted, however, unless its effect is negligible.
For recent discussions on the final-state interaction effect
see Refs. \cite{Hou}-\cite{JL2}.

In order for $\Delta_{fa}$ be nonzero it is
necessary that CP be violated and to have significant non-trivial
CP-conserving phases.  According to Eqs. (\ref{amplitude})
and (\ref{antiamplitude}),
a CP-conserving phase may arise either from the mixing
matrix $V$ or $\bar V$ or/and  from the evolution phase $e^{-i\omega_c t}$.
For a completely degenerate system, i.e.,
$\omega_a=\omega$, the evolution phase factors
\begin{eqnarray}
T_{fa}(t) &=& e^{-i\omega t} T_{fa},\label{degenerate}\\
\bar{T}_{fa}(t) &=& e^{-i\omega t} T_{fa}^*.\label{antidegenerate}
\end{eqnarray}
Although a nonreal  $T_{fa}$ implies CP violation, the contribution
to the rate difference from mixing is seen to vanish because
$|T_{fa}(t)|^2=|\bar{T}_{fa}(t)|^2$.

This result has an important implication in searching for
mechanisms for baryogenesis \cite{KT}.  It has previously
been suggested \cite{KRS} that a degenerate unstable system
might provide a resonance enhancement to the CP-violating partial
rate difference from mixing.  Our analysis shows that the outcome
would be the opposite if  the degeneracy is complete.
Instead of a resonance enhancement, we expect that
contributions to the generation
of baryon number asymmetry from unstable particle mixing is
highly suppressed whenever the particles in question have nearly
equal mass and lifetime, and is zero in the complete degenerate
limit.

One special example of a completely  degenerate system is that
the set contains only one particle,
i.e., $a=b=c=1$.    In that case
(\ref{amplitude}) and (\ref{antiamplitude}) show that particle
instability itself does not contribute to the CP-violating partial
rate difference.
The same statement applies to  systems with an arbitrary number
of unmixed particles.
In terms of perturbation theory these results imply that,
in the absence of mixing, unstable particle wave function
renormalization does not affect $\Delta_{fa}$.

Had we considered the rate difference of the eigenstates
of propagation,
i.e., $\Delta'_{f1}\equiv\Gamma(\phi'_1\to F_{f})-\Gamma(\bar{\phi}'_1\to
\bar{F}_{f})$,
we would have reached a different conclusion \cite{JL1}.
This is evident by dropping  the matrix $V$ (not $V^{-1}$) from
(\ref{amplitude}) and $\bar{V}$ from (\ref{antiamplitude}).
Phenomenologically, neglecting $V$ and $\bar V$ in (\ref{amplitude})
and (\ref{antiamplitude})
corresponds to ignoring  mixing in particle production.
As pointed out earlier \cite{Sachs1,EL},
the  propagation eigenstates
cannot be regarded as physical in the sense that
they cannot be directly produced nor detected \cite{foot1}.
As a result, $\Delta'_{fa}$ is not a physical observable.

Another important feature of Eqs. (\ref{amplitude}) and
(\ref{antiamplitude})  is that, when
the mixings are small, the diagonal mixings and phases
are irrelevant for $\Delta_{fa}$.  Indeed,
for small mixings, we expand $V$ and $\bar{V}$ as
\begin{eqnarray}
V_{ca}&=&e^{i\alpha_c}\Bigl[\delta_{ca}+\Delta V_{ca}\Bigr],\label{small}\\
\bar{V}_{ca}&=&e^{i\bar{\alpha}_c}\Bigl[
\delta_{ca} + \Delta \bar{V}_{ca}\Bigr],\label{antismall}
\end{eqnarray}
where the diagonal phases  $\alpha_c$
and $\bar{\alpha}_c$ are real.
The elements in $\Delta V$ and $\Delta\bar V$ are assumed
to be much smaller than unity.  To first order
of $\Delta V$ and $\Delta \bar{V}$ we have
\begin{eqnarray}
T_{fa}(t)&=& e^{-i\omega_a t}T_{fa} + \sum_b T_{fb}\Delta V_{ba}
              \Bigl[e^{-i\omega_b t}-e^{-i\omega_a t}\Bigr],\label{ssmall}\\
\bar{T}_{fa}(t)&=& e^{-i\omega_a t}T_{fa}^* + \sum_b
              T_{fb}^*\Delta\bar{V}_{ba}
              \Bigl[e^{-i\omega_b t}-e^{-i\omega_a t}\Bigr].\label{assmall}
\end{eqnarray}
The second terms in
Eqs. (\ref{ssmall}) and (\ref{assmall}) vanish
for diagonal mixings. Therefore,
neglecting off-diagonal mixings
the CP-conserving phase $e^{-i\omega_a t}$ factors and
$|T_{fa}(t)|^2-|\bar{T}_{fa}(t)|^2=0$.   Hence,
for small mixings the diagonal phases $\alpha_a$,
$\bar{\alpha}_a$ and $\omega_a t$ do not enter into
the determination of  $\Delta_{fa}$;
only the off-diagonal
mixings, i.e., $b\ne a$, are relevant.
This is similar to a result
of Wolfenstein's for
final-state interactions \cite{LW91}.  Thus,
for the calculation of $\Delta_{fa}$
one does not need to consider
flavor-conserving one-particle-reducible diagrams.


\section{Relations between particle and antiparticle mixings}

CPT invariance provides a relationship between $V$ and $\bar V$.
In quantum mechanics this would be obtained by studying the Hamiltonian.
The existence of the `standard model' underlines the
usefulness of a field theoretical analysis.
In field theory the relation between $V$ and $\bar V$  can be easily
obtained from
particle propagator.
Consider situations in which $\phi_a$ is a scalar.
At tree level, the propagator of $\phi_a$ is
\begin{eqnarray}
i\tilde{\Delta}_{ba}^{(0)}(P^2)=i\Bigl[P^2-m_a^{(0) 2}\Bigr]^{-1}
\delta_{ba},
\label{barepropagator}
\end{eqnarray}
where $m_a^{(0)}$ is the bare mass of $\phi_a$, and
we have implicitly assumed that
$|\phi_a\rangle$ is an eigenstate of zeroth order of
$H_{\rm weak}$.
Including one-loop
corrections  the following changes occur in
(\ref{barepropagator})
\begin{eqnarray}
m_{a}^{(0) 2}\delta_{ba}\to {\cal{M}}^2_{ba}.\label{replacment}
\end{eqnarray}
where  ${\cal{M}}^2$ is the square
of an effective complex mass matrix
\begin{eqnarray}
{\cal{M}}^2_{ba}=\tilde{{\cal{M}}}_{ba}^2 - i
\tilde{\Gamma}_{ba}^2,
\label{decaymatrix}
\end{eqnarray}
in which $\tilde{{\cal{M}}}$ is the effective mass matrix  and
$\tilde{\Gamma}^2={1\over 2}(\tilde{\cal{M}}\Gamma+\Gamma\tilde{\cal{M}})$,
where $\Gamma$ is the decay matrix, and $\Gamma^2/4$ has
been neglected.
Both $\tilde{{\cal{M}}}$ and $\tilde{\Gamma}^2$ are hermitian,
but ${\cal{M}}^2$ is not.

For a given $P^2$,
${\cal{M}}^2$ can be diagonalized  by a transformation
\begin{eqnarray}
\Bigl[Q{\cal{M}}^2Q^{-1}\Bigr]_{ba}=(m_a^2-im_a\gamma_a
)\delta_{ba},
\label{diagonal}
\end{eqnarray}
where $Q$ is a complex matrix.
The one-loop regularized propagator can be written as
\begin{eqnarray}
i\tilde{\Delta}_{ba}^{(1)}(P^2)=iQ_{bc}^{-1}
\Bigl[P^2-m_c^2+im_c\gamma_c\Bigr]^{-1}Q_{ca},\label{oneloop-prop}
\end{eqnarray}
where for simplicity we have neglected terms of order $\gamma^2$.
Evidently, $i\Bigl[P^2-m_c^2+im_c\gamma_c\Bigr]^{-1}$ is the
propagator of $\phi_c'$.  It follows that
\begin{eqnarray}
&&V_{ba}=Q_{ba},\label{Oba}\\
&&\bar{V}_{ba}=Q^{-1}_{ab},\label{Obarab}
\end{eqnarray}
and thus the relationship between $\bar{V}$ and $V$ is
\begin{eqnarray}
\bar{V}_{ba}=V^{-1}_{ab}.\label{relation}
\end{eqnarray}
For  stable particles ${\cal{M}}^2$ is hermitian,
$Q$ is unitary and hence so are
$V$ and $\bar V$, and (\ref{relation}) reduces
to the known result $\bar V=V^*$.
Substituting (\ref{relation}) into (\ref{antiamplitude})  one sees that
the time-dependent mixing matrix in the antiparticle decay is
$\sum_cV_{ac}^{-1}e^{-i\omega_c t}V_{cb}$, which differs from that
in the particle decay (\ref{amplitude}) by exchanging
the indices $a$ and $b$ (a consequence of time reversal).

It is important that ${\cal{M}}^2$ is
momentum dependent.  This is  necessary if the orthonormality conditions
are to be maintained for both $|\phi_a\rangle$ and $|\phi_a'\rangle$
 \cite{foot2}.
In practice this does not introduce any additional
complication, as
$P^2$ is always fixed by the on-shell condition
once the initial state is specified.

We now focus on a case of special interest,  small mixing.  By small
mixing we mean that (1) the width differences of the particles
are much smaller than their mass differences and (2)
the off-diagonal elements in ${\cal{M}}^2$ can be treated as
a perturbation.  In that case $\Delta V$ and $\Delta\bar V$
defined by (\ref{small}) and (\ref{antismall}),
may be separated into their dispersive and absorptive parts
\begin{eqnarray}
&&\Delta V_{ba}= \Delta V_{ba}^{(D)} + i  \Delta V_{ba}^{(A)},\label{DVDA}\\
&&\Delta \bar{V}_{ba}
= \Delta \bar{V}_{ba}^{(D)} + i  \Delta \bar{V}_{ba}^{(A)}.\label{DVbarDA}
\end{eqnarray}
Phenomenologically,
$\Delta V_{ba}^{(D)}$ and $\Delta V_{ba}^{(A)}$
correspond to the mixings
arising from the mass and decay matrices, respectively.  The hermiticity
of the mass and decay matrices then implies
\begin{eqnarray}
&& \Delta V_{ba}^{(D,A)}=-\Delta V_{ab}^{(D,A) *},\label{DVhermiticity}\\
&&  \Delta \bar{V}_{ba}^{(D,A)}
    =-\Delta \bar{V}_{ab}^{(D,A) *}.\label{DVbarhermiticity}
\end{eqnarray}
The solution to (\ref{relation}) satisfying the constraints of
(\ref{DVhermiticity}) and (\ref{DVbarhermiticity}) is
\begin{eqnarray}
\Delta V_{ba}^{(D,A)}=\Delta\bar{V}_{ba}^{(D,A) *}.\label{srelation}
\end{eqnarray}
This is a much simplified version of (\ref{relation}),
valid for small mixings.

With (\ref{srelation}) one can have a simple expression for
$\Delta_{fa}$.  From
(\ref{ssmall}), (\ref{assmall}) and (\ref{srelation}) we find
that the time-differentiated CP-violating partial rate difference
is
\begin{eqnarray}
|T_{fa}(t)|^2-|\bar{T}_{fa}(t)|^2=4\sum_b\rho_b
\Bigl\{\sin\beta_b e^{2{\rm Im}\omega_a t}
-\sin[{\rm Re}(\omega_a-\omega_b)t + \beta_b]e^{{\rm Im}(\omega_a+\omega_b)t}
\Bigr\},
\label{smallresult}
\end{eqnarray}
where
\begin{eqnarray}
&& \rho_b=\sqrt{\Bigl[{\rm Im}\Bigl(\Delta V_{ba}^{(A)}T_{fa}^*T_{fb}
             \Bigr)\Bigr]^2
             +\Bigl[{\rm Im}\Bigl(\Delta V_{ba}^{(D)}T_{fa}^*T_{fb}
             \Bigr)\Bigr]^2},
\label{modular}\\
&& \tan\beta_b={\rm Im}\Bigl[\Delta V_{ba}^{(A)}T_{fa}^*T_{fb}\Bigr]/
              {\rm Im}\Bigl[\Delta V_{ba}^{(D)}T_{fa}^*T_{fb}\Bigr].
\label{phasshift}
\end{eqnarray}
In the absence of CP violation $\Delta V^{(D,A)}$ and $T_{fa}$ are
real, $\rho_b=0$ and hence Eq. (\ref{smallresult}) vanishes.
As pointed out earlier,
Eq. (\ref{smallresult}) shows explicitly that
for diagonal mixing, i.e., $b=a$, the phase makes no contribution.
If the mass and decay matrices commute, $
\Delta V^{(A)}=0$ and   (\ref{smallresult})
reduces to
\begin{eqnarray}
|T_{fa}(t)|^2-|\bar{T}_{fa}(t)|^2=-4\sum_b
{\rm Im}\Bigl[\Delta V_{ba}^{(D)}T_{fa}^*T_{fb}\Bigr]
\sin\Bigl[{\rm Re}(\omega_a-\omega_b)t\Bigr]e^{{\rm Im}(\omega_a+\omega_b)t}
{}.
\label{commutesmallresult}
\end{eqnarray}
The result given by
(\ref{commutesmallresult})
can also  be
obtained from a general
formalism developed recently by Gronau and Rosner
\cite{GR} for a $2\times 2$ system.

Within our approximation
all oscillatory terms are integrated to zero.  Hence
\begin{eqnarray}
\Delta_{fa} & = &
\int d\Omega\int^{\infty}_{0}{dt\over \tau_a}
    \Bigl[ |T_{fa}(t)|^2-|\bar{T}_{fa}(t)|^2\Bigr]\nonumber\\
& = &
4\int d\Omega \sum_{b\ne a}{\rm Im}\Bigl[\Delta V_{ba}^{(A)}
     T_{fa}^*T_{fb}\Bigr],\label{ssmallresult}
\end{eqnarray}
where $1/\tau_a=- 2{\rm Im}\omega_a$ is the
width of the particle and $\int d\Omega$
 represents a phase space sum.
Eq. (\ref{ssmallresult}) shows that for the
calculation of $\Delta_{fa}$ one only
needs to consider off-diagonal mixings in the decay matrix.

For small mixings it is easy to show that
\begin{eqnarray}
\sum_f\Delta_{fa}=0.\label{CPT}
\end{eqnarray}
This relation follows from unitarity of $S$-matrix,  which in
the present situation implies
\begin{eqnarray}
\sum_{f}T_{fb}^*T_{fa}|_{P^2=m_a^2}\propto
\tilde{\Gamma}_{ba}^2\propto \Delta V^{(A)}_{ba}.
\label{unitarity}
\end{eqnarray}
Thus, $\sum_{f,b\ne a}\Delta V_{ab}^{(A)}T_{af}^*T_{bf}$
is real and Eq. (\ref{CPT}) follows as a consequence.
Since the final-state interaction contribution to the total
rate difference of a particle and its antiparticle is known to
vanish \cite{LW91}, the result of (\ref{CPT})
is in accordance with the usual expectation that a particle
and its antiparticle have the same lifetime \cite{lifetime}.

\section{An illustrative  example for small mixing}

In this section we  show how to apply (\ref{ssmallresult})
to model calculations.
Consider a system containing two heavy scalars $S_{a,\alpha}\ (a=1,2)$
and $\alpha$ as a color index.
The lowest order interaction of the system is given by
\begin{eqnarray}
{\cal{L}}_{I}=
             G_{1a}\bar u^c_{R,\alpha}d_{R,\beta}S_{a,\gamma}
              \epsilon^{\alpha\beta\gamma}
             +G_{2a}\bar u_{R,\alpha}e^c_{R}S_{a,\alpha}
             +h.c.,\label{lagrangian}
\end{eqnarray}
where  $u,  d$ and $e^c$ are the
charged fermions of the first
generation, which are
considered as massless.
This interaction
can arise from an $SU(5)$ GUT theory with two 5-plets of Higgs fields.
Since the two final states into which the heavy
scalars can decay have different baryon number, (\ref{lagrangian})
provides an interaction for baryogenesis via heavy scalar decays.

For convenience, we choose as a basis $S_{a,\alpha}$ the
mass eigenstate fields and select
$|F_1\rangle=|\bar{d}_Ru_R^c\rangle$ and $|F_2\rangle=|e_R\bar{u}_R^c\rangle$.
The lowest order transition matrix for
 $|S_{1,2}\rangle\to |F_{1,2}\rangle$
is
\begin{eqnarray}
T= C
\left(\begin{array}{cc} \sqrt 2G_{11}   &\sqrt 2 G_{12}\\
                        G_{21}          & G_{22}\end{array}\right),
\label{T}
\end{eqnarray}
where $C$ is an over all  normalization constant.  The factor
$\sqrt 2$ in $T_{11}$ and $T_{12}$ is introduced to account for
the effect of summing over the indices of $\epsilon^{\alpha\beta\gamma}$
in calculating the squares of the elements.

To determine $\Delta V^{(A)}$ we
 consider corrections
 to the scalar propagators to second order of ${\cal{L}}_I$.
A simple calculation shows that
the regularized
 elements in the  complex-mass-matrix square
(\ref{decaymatrix})  are
(remember $\tilde{M}^2$ and $\tilde{\Gamma}^2$ are momentum
dependent)
\begin{eqnarray}
&& \tilde{M}_{ba}^2=m_{a}^{(0)2}\delta_{ba} -  {P^{2}\over 16\pi^2}
 \Bigl[2G_{1b}^*G_{1a}+G_{2b}^*G_{2a}\Bigr]
   \Bigl[{2\over 4-n}+{3\over 2}-\gamma_E-\ln{P^2\over 4\pi\mu_0^2}\Bigr],
\label{regulmass}\\
&& \tilde{\Gamma}_{ba}^2= {P^2\over 16\pi}
\Bigl[2G_{1b}^*G_{1a}+G_{2b}^*G_{2a}\Bigr],\label{regulwidth}
\end{eqnarray}
where $n$ is the dimension of regularization, $\mu_0^2$ is
the ultraviolet cut-off  and
$\gamma_E$ is Euler's number.
Following the standard technique \cite{LOY} we find that
the matrix which  diagonalizes ${\cal{M}}^2$
according to (\ref{diagonal}) is
\begin{eqnarray}
V=
\left(\begin{array}{cc} \cos\theta  &-\sin\theta \\
                       \sin\theta   &\cos\theta\end{array}\right)
\left(\begin{array}{cc} e^{-i\delta}  & 0\\
                        0             & e^{i\delta}\end{array}\right),
\label{Vmixing}
\end{eqnarray}
where $\theta={\rm Re}\theta + i{\rm Im}\theta$ is complex
\begin{eqnarray}
\tan 2\theta =  2 {\hat{M}^{2}+i|\tilde{\Gamma}_{12}^{2}|\over
         (\tilde{M}_{11}^2-\tilde{M}_{22}^2)
         -i(\tilde{\Gamma}_{11}^2-\tilde{\Gamma}_{22}^2)},\label{cangle}
\end{eqnarray}
with the real quantity
$\hat{M}^2=(P^2/16\pi^2)|2G_{11}^*G_{12}+G_{21}^*G_{22}|[
{2\over 4-n} + {3\over 2} -\gamma_E -\ln(P^2/4\pi\mu_0^2)]$.
The phase $\delta$ is real and determined by
\begin{eqnarray}
\delta={1\over 2} \arg \Bigl[2G_{11}^*G_{12}+G_{21}^*G_{22}\Bigr].
\label{phasese}
\end{eqnarray}
The eigenvalues are
\begin{eqnarray}
m_{1,2}^2-im_{1,2}\gamma_{1,2}= &&{1\over 2}\Bigl\{
       (\tilde{M}_{11}^2+\tilde{M}_{22}^2)
       -i(\tilde{\Gamma}_{11}^2+\tilde{\Gamma}_{22}^2)\nonumber\\
&& \pm
       \sqrt{[(\tilde{M}_{11}^2-\tilde{M}_{22}^2)
       -i(\tilde{\Gamma}_{11}^2-\tilde{\Gamma}_{22}^2)]^2
             + 4[\hat{M}^{2} + i|\tilde{\Gamma}_{12}^{2}|]^2}\Bigr\}.
\label{eigenvalue}
\end{eqnarray}

In the small mixing limit, i.e.,
\begin{eqnarray}
|\tilde{M}_{11}^2-\tilde{M}_{22}^2|\gg
|\tilde{\Gamma}^2_{11}-\tilde{\Gamma}^2_{22}|,
|\tilde{M}_{12}^{2}|, |\tilde{\Gamma}_{12}^{2}|,
\label{smallcondition}
\end{eqnarray}
one has from (\ref{Vmixing}) that
\begin{eqnarray}
\Delta V^{(A)}=
\left(\begin{array}{cc} 0  &-{\rm Im}\theta e^{i2\delta}\\
             {\rm Im}\theta e^{-i2\delta}  &0\end{array}\right),
\label{smallVmixing}
\end{eqnarray}
where
\begin{eqnarray}
{\rm Im}\theta e^{i2\delta}=\Bigl[{\rm Im}\theta e^{-i2\delta}\Bigr]^*
={P^2[2G_{11}^*G_{12}+G_{21}^*G_{22}]
\over 16\pi [\tilde{M}_{11}^2-\tilde{M}_{22}^2]}.
\label{Imangle}
\end{eqnarray}
Substituting (\ref{T}) and (\ref{smallVmixing})  into
(\ref{ssmallresult}) and making use of
the relation  $m_{1,2}^{(0)2}=\tilde{M}_{11,22}^2=m_{1,2}^2$,
which is valid to
zeroth order in ${\cal{L}}_I$,
we obtain for $S_1\to F_{1,2}$, in which
$P^2=m_1^2$,
\begin{eqnarray}
\Delta_{11}=
-\Delta_{21}
={\Omega_1\over 2\pi}
{m_1^2\over (m_1^2-m_2^2)}{\rm Im}
\Bigl[G_{11}^*G_{12}G_{21}G_{22}^*\Bigr],
\label{delta11}
\end{eqnarray}
where $\Omega_1=m_1/16\pi$ is a phase space factor for $S_1$.
Also, for the $S_2$ decays we have  $P^2=m_2^2$ and
\begin{eqnarray}
\Delta_{12}=
-\Delta_{22}
={\Omega_2\over 2\pi}
{m_2^2\over (m_2^2-m_1^2)}{\rm Im}
\Bigl[G_{11}G_{12}^*G_{21}^*G_{22}\Bigr],\label{delta12}
\end{eqnarray}
where $\Omega_2=m_2/16\pi$.
These results are in complete
agreement with those obtained from a direct Feynman diagram calculation
\cite{LS1}.  Contributions to $\Delta_{fa}$ from vertex corrections
can be found in \cite{NW,BR,LS1}.

\bigskip

\section{An illustrative example for large mixing}

In the small mixing limit (\ref{smallcondition}) one can simply
use the regularized (but not renormalized) formalism to
compute $\Delta_{fa}$.  The unphysical quantities in $\tilde{M}^2$
do not enter.  However, to go beyond this limit we must introduce
a renormalization prescription to remove the divergences.
A pedagogic introduction illustrating how this may be done for
the example discussed above is given in Appendix A.

{}From a phenomenological viewpoint,
the mixing phenomena under consideration is determined by
the renormalized interaction lagrangian \cite{foot4}
\begin{eqnarray}
 {\cal{L}}_I=
g_{1a}\bar u^c_{R,\alpha}d_{R,\beta}S_{a,\gamma}
 \epsilon^{\alpha\beta\gamma}
             +g_{2a}\bar u_{R,\alpha}e^c_{R}S_{a,\alpha}
             +h.c.,\label{Rlagrangian}
\end{eqnarray}
where $g_{fa}\ (f=1,2)$ are the renormalized couplings in
the weak eigenstate basis.  The square of a renormalized complex mass matrix
is
\begin{eqnarray}
{\cal{M}}^2=
\left(\begin{array}{cc} \tilde{M}_{11}^2   &  \tilde{M}_{12}^2\\
                            \tilde{M}_{12}^{2*} & \tilde{M}_{22}^2
\end{array}\right)
-i
\left(\begin{array}{cc} \tilde{\Gamma}_{11}^2 & \tilde{\Gamma}_{12}^2\\
             \tilde{\Gamma}_{12}^{2*}      &   \tilde{\Gamma}_{22}^2
\end{array}\right),\label{RCmassmatrix}
\end{eqnarray}
where
 $\tilde{M}^{2}_{ba}$ and $g_{fa}$  are
the parameters of the model
determined experimentally (at some
scale).
The parameters in $\tilde{\Gamma}^2_{ba}$  are calculable
\begin{eqnarray}
\tilde{\Gamma}_{ba}^2={P^2\over 16\pi^2}\Bigl[
     2g_{1b}^*g_{1a} + g_{2b}^*g_{2a}\Bigr].\label{Rdecaymatrix}
\end{eqnarray}

A convenient form  for the matrix which diagonalizes (\ref{RCmassmatrix})
according to (\ref{diagonal}) is
\begin{eqnarray}
V=\left(\begin{array}{cc} \cos\theta    & -\sin\theta e^{i\phi}\\
              \sin\theta e^{-i\phi}     &  \cos\theta\end{array}\right).
\label{RV}
\end{eqnarray}
Again, the mixing angle $\theta$ is complex
\begin{eqnarray}
\tan^22\theta=4{\Bigl[\tilde{M}^2_{12}-i\tilde{\Gamma}^2_{12}\Bigr]
                \Bigl[\tilde{M}^{2*}_{12}-i\tilde{\Gamma}^{2*}_{12}\Bigr]
                \over
                \Bigl[\Bigl(\tilde{M}^2_{11}-\tilde{M}^2_{22}\Bigr)
                     -i\Bigl(\tilde{\Gamma}_{11}^2-
                             \tilde{\Gamma}_{22}^2\Bigr)\Bigr]^2}.
\label{Rangle}
\end{eqnarray}
The phase $\phi$ is also complex determined by
\begin{eqnarray}
e^{-i2\phi}={\tilde{M}^{2*}_{12}-i\tilde{\Gamma}^{2*}_{12}\over
             \tilde{M}^{2}_{12}-i\tilde{\Gamma}^{2}_{12}}.\label{Rphase}
\end{eqnarray}
The eigenvalues are
\begin{eqnarray}
\omega_{1,2}^2 & =&
\Bigl[m_{1,2}  -{i\over 2}\gamma_{1,2}\Bigr]^2\nonumber\\
&=&
{1\over 2}\Bigl[ \Bigl(\tilde{M}^2_{11}+\tilde{M}^2_{22}\Bigr)
                 -i\Bigl(\tilde{\Gamma}^2_{11} +
                 \tilde{\Gamma}^2_{22}\Bigr)\Bigr]\nonumber\\
&\pm {1\over 2}&\sqrt{\Bigl[\Bigl(\tilde{M}^2_{11}-\tilde{M}^2_{22}\Bigr)
                  -i\Bigl(\tilde{\Gamma}_{11}^2-\tilde{\Gamma}_{22}^2\Bigr)
                  \Bigr]^2
               + 4\Bigl(\tilde{M}^2_{12}-i\tilde{\Gamma}^2_{12}\Bigr)
                  \Bigl(\tilde{M}^{2*}_{12}-i\tilde{\Gamma}^{2*}_{12}\Bigr)}
          .\label{Reigenvalue}
\end{eqnarray}

If $\arg \tilde{M}^2_{12}=\arg \tilde{\Gamma}_{12}^2$,
$\phi=2\arg  \tilde{M}^2_{12}= 2\delta$ is real.
Compared to the regularized formula (\ref{Vmixing}),
one sees that Eq. (\ref{RV}) differs from
(\ref{Vmixing}) only  by a diagonal phase matrix.  This difference
has no physical significance.  The diagonal phase matrix can
be removed by a suitable choice of phase convention.

Let us now turn back to $\Delta_{fa}$. Eqs. (\ref{Rdecaymatrix}) to
(\ref{Reigenvalue}) enable us to compute $\Delta_{fa}$
for arbitrary mixing angle and phase.
The results are only limited by the validity of
perturbation expansion of ${\cal{L}}_I$. Here one
should use the general formulas (\ref{amplitude}),
(\ref{antiamplitude}) and (\ref{relation}).
In applying Eqs. (\ref{RV}) to (\ref{Reigenvalue})
one should also be very careful whenever the parameters
are near the branch cuts in the complex parameter space.

As an illustration we consider that
the initial (renormalized) state is an eigenstate of mass
matrix, i.e., $\tilde{M}_{12}^2\to 0$, but not an eigenstate
of the decay matrix.  In that case $\phi$ is real and
\begin{eqnarray}
\sin 2\theta e^{-i\phi}=i{P^2\over 8\pi |\omega_1^2-\omega_2^2|^2}
(\omega_1^{2*}-\omega_2^{2*})
\Bigl[2g_{11}g_{12}^*+g_{21}g_{22}^*\Bigr].\label{Lmixing}
\end{eqnarray}
We find
\begin{eqnarray}
&&\Delta_{11} = -\Delta_{21}\nonumber\\
&&={\Omega_1 m_1^2{\rm Im}
\Bigl[g_{11}^*g_{12}g_{21}g_{22}^*\Bigr]
\over 2\pi |\omega_1^2-\omega_2^2|^2}\int_0^{\infty}{dt\over \tau_1}
{\rm Re}\Bigl[\Bigl(\omega_1^2-\omega_2^2\Bigr)\Bigl(
c^2 e^{-i\omega_1t} + s^2 e^{-i\omega_2t}\Bigr)
\Bigl(e^{i\omega_1^*t}-e^{i\omega_2^*t}\Bigr)\Bigr],
\label{Ldelta11}
\end{eqnarray}
where $c=\cos\theta$ and $s=\sin\theta$.
Also,
\begin{eqnarray}
&&\Delta_{12} = -\Delta_{22}\nonumber\\
&&={\Omega_2 m_2^2 {\rm Im}
\Bigl[g_{11}g_{12}^*g_{21}^*g_{22}\Bigr]
\over  2\pi |\omega_2^2-\omega_1^2|^2}\int_0^{\infty}{dt\over \tau_2}
{\rm Re}\Bigl[\Bigl(\omega_2^2-\omega_1^2\Bigr)\Bigl(
c^2 e^{-i\omega_2t} + s^2 e^{-i\omega_1t}\Bigr)
\Bigl(e^{i\omega_2^*t}-e^{i\omega_1^*t}\Bigr)\Bigr].
\label{Ldelta12}
\end{eqnarray}
In (\ref{Ldelta11}) and (\ref{Ldelta12}) the mixing angle
$\theta$ is arbitrary.

In the limit in which the mixing angle is small due to
$|\tilde{\Gamma}_{11}^2-\tilde{\Gamma}_{22}^2|\gg
|\tilde{M}_{11}^2-\tilde{M}_{22}^2|, |\tilde{\Gamma}_{12}^2|$,
Eqs. (\ref{Ldelta11}) and (\ref{Ldelta12}) are simplified to
\begin{eqnarray}
\Delta_{11} &=&-\Delta_{21}= {\Omega_1\over 2\pi}{m_1^2\over m_1^2-m_2^2}
   {\rm Im}\Bigl[g_{11}^*g_{12}g_{21}g_{22}^*\Bigr]
\Bigl[{m_1^2-m_2^2\over \tilde{\Gamma}_{11}^2-\tilde{\Gamma}_{22}^2}
\Bigr]^2,\label{smdecay11}\\
\Delta_{12}&=&-\Delta_{22}= {\Omega_2\over 2\pi}{m_2^2\over m_2^2-m_1^2}
   {\rm Im}\Bigl[g_{11}g_{12}^*g_{21}^*g_{22}\Bigr]
\Bigl[{m_1^2-m_2^2\over \tilde{\Gamma}_{11}^2-\tilde{\Gamma}_{22}^2}
\Bigr]^2.\label{smdecay12}
\end{eqnarray}
Compared to (\ref{delta11}) and (\ref{delta12}) (
after renormalization $G_{fa}$ are replaced by $g_{fa}$),
these results are further suppressed by a
ratio $(m_1^2-m_2^2)^2/ (\tilde{\Gamma}_{11}^2-\tilde{\Gamma}_{22}^2)^2\ll 1$.
In particular, all the CP-violating partial rate differences vanish
in the limit $m_1=m_2$.
This is in contrast to what one might
have expected, based upon a naive extrapolation from (\ref{delta11})
and (\ref{delta12}).
 Since
$\Delta_{fa}$ depends on the imaginary part of the mixing angle
(\ref{ssmallresult}),  it must vanish
whenever ${\rm Im}\theta=0$.  For the case at hand,
$\theta$ is real when $\tilde{M}_{12}^2=0$ and $m_1=m_2$.


\section{Conclusion}

We have discussed a formalism for  unstable
particle mixing base upon one-loop renormalization of
field theory, with emphasize on its applications to
CP-violating physical processes.  Among various interesting
results,  we have found that, for a completely degenerate system,
unstable particle mixing does not contribute to  the CP-violating
partial rate difference.  In particular, in the absence of
mixing, unstable particle wave function renormalization does
not introduce any additional effect for the CP asymmetry.
When the mixing is small we show that only the off-diagonal
mixings and phases are relevant for CP violation.

We have used a simple example to show   how to apply the formalisms developed
in this paper in model calculations
for arbitrary mixing.  The basic steps for
renormalizing unstable particle mixing
are also outlined.

\bigskip

{\bf Acknowdgement}

We gratefully acknowledge discussions with D. Chang and Paul Langacker.
This work was supported in part by
an SSC fellowship (J.L.) from the Texas National
Research Laboratory Commission and by  the U. S. Department of Energy
under contract DOE AC02-76-ERO-3071.
\newpage

\centerline{\large\bf {Appendix A}}

\bigskip

It is most convenient to introduce a renormalization
prescription in the weak eigenstate basis. We parameterize
the interaction lagrangian
as
\begin{eqnarray}
{\cal{L}}_{I}=
             g^{(0)}_{1a}\bar u^c_{R,\alpha}d_{R,\beta}S^{(0)}_{a,\gamma}
              \epsilon^{\alpha\beta\gamma}
             +g^{(0)}_{2a}\bar u_{R,\alpha}e^c_{R}S^{(0)}_{a,\alpha}
             +h.c.,\label{barelagrangian}
\end{eqnarray}
where $g_{fa}^{(0)}\ (f=1,2)$ are the weak eigenstate bare couplings.
The one-loop regularized inverse propagator can then be
written as
\begin{eqnarray}
\tilde{\Delta}_{ba}^{-1}(P^2,\mu^2)&=&P^2\delta_{ba} - m_{ba}^{(0)2} +
       \Sigma_{ba}^{R}(P^2,\mu^2)\nonumber\\
&+& {1\over 16\pi^2}
\Bigl[2g_{1b}^*g_{1a} + g_{2b}^*g_{2a}\Bigr]
\Bigl[P^2\Bigl(f_0-1-\ln{\mu^2\over \mu_0^2}\Bigr)
   + \mu^2 + i\pi P^2\theta(P^2)\Bigr],\label{oneloopinverse}
\end{eqnarray}
where $m_{ba}^{(0) 2}$ is the square of the bare mass matrix,
$f_0={2\over 4-n}+{3\over 2} - \gamma_E+\ln 4\pi$,
$g_{fa}\ (f=1,2)$ is the renormalized couplings (to the lowest order
$g_{fa}=g_{fa}^{(0)}$) and $\theta$ is the
$\theta$-function.  $\Sigma_{ba}^{R}(P^2,\mu^2)$
is the renormalized self-energy
\begin{eqnarray}
\Sigma_{ba}^{R}(P^2,\mu^2)=
{1\over 16\pi^2}\Bigl[2g_{1b}^*g_{1a} + g_{2b}^*g_{2a}\Bigr]
\Bigl[(P^2-\mu^2)+P^2\ln{P^2\over \mu^2} + i\pi P^2\Big(\theta(P^2)
-\theta(\mu^2)\Bigr)\Bigr],\label{Rself}
\end{eqnarray}
which has the standard features
\begin{eqnarray}
&&\Sigma^{R}_{ba}(\mu^2,\mu^2)=0,\label{normal1}\\
&&{\partial\over\partial P^2}\Sigma^{R}_{ba}(P^2,\mu^2)\vert_{P^2=\mu^2}
=0,\label{normal2}
\end{eqnarray}
and $\mu^2$ is the subtraction point.  In practice it
is convenient to choose $\mu^2$ be the invariant mass square of the
initial particle.

Following the standard method the renormalized inverse
propagator is
\begin{eqnarray}
\tilde{\Delta}_{R,ba}^{-1}(P^2,\mu^2)=
P^2\delta_{ba} -\tilde{M}^2_{ba}(\mu^2) + i\tilde{\Gamma}^2_{ba}(P^2)
+\Sigma_{ba}^{R}(P^2,\mu^2),\label{renormalinverse}
\end{eqnarray}
where
$\tilde{\Gamma}^2_{ba}(P^2)$
is  given by (\ref{Rdecaymatrix}) and
\begin{eqnarray}
\tilde{M}_{ba}^2(\mu^2)&=&
m_{ba}^{(0) 2}
 -{\mu^2\over 16\pi^2}\Bigl[2g_{1b}^*g_{1a} + g_{2b}^*g_{2a}\Bigr]\nonumber\\
&-&{1\over 16\pi^2}\sum_c\Bigl\{\Bigl[2g_{1b}^*g_{1c} + g_{2b}^*g_{2c}\Big]
\tilde{M}^2_{ca} + \tilde{M}_{bc}^2
\Bigl[2g_{1c}^*g_{1a} + g_{2c}^*g_{2a}\Big]\Bigr\}
\Bigl[f_0-1-\ln{\mu^2\over \mu_0^2}\Bigr]\label{Rmassmatrix}
\end{eqnarray}
is the square of the renormalized mass matrix.   The renormalized
weak eigenstate fields are related to the bare
fields by
\begin{eqnarray}
S_{a,\alpha}^{(0)}=\sum_b\Bigl\{
\delta_{ba}-{1\over 16\pi^2}\Bigl[2g_{1b}^*g_{1a} + g_{2b}^*g_{2a}\Bigr]
\Bigl[f_0-1-\ln{\mu^2\over \mu_0^2}\Bigr]\Bigr\}S_{b,\alpha}.\label{Rwave}
\end{eqnarray}
The divergencies in the interaction lagrangian associated
with wave function renormalization and vertex corrections
are finally absorbed into coupling
constant renormalization, i.e.,
$g_{fa}^{(0)}\to g_{fa}$.

Once the renormalized parameters are determined
at a given scale $\mu^2$ by experiments,
the separation of the divergent term $f_0$ from $P^2$ and $\mu^2$ in
(\ref{Rself}),
(\ref{Rmassmatrix}) and (\ref{Rwave}) allows us to predict
these parameters at any other scale.

\bigskip


\centerline{\large\bf Appendix B}

\bigskip

In this appendix we outline the basic ideas of diagonalizing
an arbitrary $n\times n $ complex matrix ${\cal{H}}$.
${\cal{H}}$ can always be diagonalized by a bi-unitary transformation, i.e.,
$\Bigl[V_L{\cal{H}}V_R^{\dag}\Bigr]_{ba}=\lambda_b\delta_{ba}$,
where $V_L$ and $V_R$ are unitary and the eigenvalues $\lambda_a$
are real. However, this procedure is not useful for us unless
$V_LV_R^{\dag}=1$; otherwise
the kinetic energy part of the lagrangian will not remain
diagonal under such a transformation.

Methods for diagonalizing a $2\times 2$
complex nonhermitian matrix are known \cite{LOY}.  We have employed one of
them in our analysis in section 5.  Here we
extend them to arbitrary cases.  Consider an arbitrary
$n\times n$ complex matrix ${\cal{H}}$.  In general ${\cal{H}}$
has $2n^2$ parameters.  It is convenient to separate ${\cal{H}}$
in to a hermitian and an antihermitian part
\begin{eqnarray}
{\cal{H}}=M-i\Gamma,\label{B1}
\end{eqnarray}
where
\begin{eqnarray}
&& M= M^{\dag}= {{\cal{H}} + {\cal{H}}^{\dag}\over 2},\label{B2}\\
&& \Gamma=\Gamma^{\dag}
              =i {{\cal{H}} - {\cal{H}}^{\dag}\over 2}.\label{B3}
\end{eqnarray}
In the study of unstable particle mixing $M$ and $\Gamma$
are the mass and decay matrices, respectively.

Let us start from  simple cases involving real $M$ and $\Gamma$.
These correspond to situations in which the mixed
unstable-particle system conserves CP.
Now, ${\cal{H}}$ is symmetric.  It has
$n(n+1)$ independent parameters.  We can diagonalize such
a ${\cal{H}}$
by an orthogonal transformation \cite{Langacker}
\begin{eqnarray}
O{\cal{H}}O^T=
\left(\begin{array}{ccc} \lambda_1&\          & 0\\
                        \        & \ddots    & \ \\
                        0        &\          & \lambda_n\end{array}\right),
\label{B4}
\end{eqnarray}
where
$OO^T=1$.
$O$ differs from a usual orthogonal matrix
by having complex mixing angles, i.e., $\theta_i={\rm Re}\theta_i
+{\rm Im}\theta_i\ (i=1,...,n(n-1)/2)$.
It should be emphasized that a complex mixing angle
is not equivalent to a real mixing angle times a real phase.
Indeed,  $|\cos\theta|$ could be bigger than $1$
if $\theta$ is complex, whereas $|\cos\theta e^{i\alpha}|\leq 1$
if both $\theta$ and $\alpha$ are real.
Nevertheless, trigonometric functions with complex angles
have very similar properties, such as $\sin^2\theta + \cos^2\theta =1$ etc.,
as the elementary trigonometric functions.
This is the advantage of introducing
complex mixing angles.

The $n(n+1)$ parameters of ${\cal{H}}$ determine
the $n$ complex eigenvalues, which have $2n$  parameters, and
the $n(n-1)/2$ complex angles involving  $n(n-1)$ parameters.

This idea can be extended to situations in which both $M$ and $\Gamma$
are complex, provided the phases involved in the diagonalization
are chosen as complex as well \cite{DC}.

Let us first review the diagonalization of a hermitian matrix $H$
\begin{eqnarray}
UHU^{-1}= \left(\begin{array}{ccc} \lambda_1&\  & 0\\
                        \  & \ddots & \ \\
                        0  &\     & \lambda_n\end{array}\right),
\label{B5}
\end{eqnarray}
where $\lambda_i$ are real. $U^{-1}=U^{\dag}$ is unitary with
$n(n-1)/2$ real angles and $n(n+1)/2$ real phases.  However,
not all phases of $U$ are determined by (\ref{B5}). To be specific,
we write
\begin{eqnarray}
U=KU',\label{B6}
\end{eqnarray}
where $U'$ is a {\it reduced} unitary matrix with $n(n-1)/2$
angles and $n(n-1)/2$ phase, $K$ is diagonal with $n$ phases
\begin{eqnarray}
K=\left(\begin{array}{ccc} e^{i\alpha_1} &\  & 0\\
            \ & \ddots & \ \\
            0 & \  & e^{i\alpha_n}\end{array}\right).
\label{B7}
\end{eqnarray}
Clearly, if $U$ diagonalizes $H$ so does its reduced matrix $U'$.

An arbitrary ${\cal{H}}$ has $2n^2$ parameters: among them, $n^2$
in $M$ and the rest in $\Gamma$.  The transformation that
leaves the unity matrix invariant and diagonalizes ${\cal{H}}$
is
\begin{eqnarray}
V{\cal{H}} V^{-1}=\left(\begin{array}{ccc} \lambda_1&\  & 0\\
                        \  & \ddots & \ \\
                        0  &\     & \lambda_n\end{array}\right),
\label{B8}
\end{eqnarray}
where $\lambda_i$ are complex.  In practice we may parameterize
$V$ in terms of a reduced unitary matrix, such as $U'$ discussed above,
but change all the mixing angles and phases into complex variables.
Now, the $2n^2$ parameters determine the $n$ complex eigenvalues
which involve $2n$ parameters, the $n(n-1)/2$ complex mixing angles
(with $n(n-1)$ parameters) and the $n(n-1)/2$ complex phases,
which also have $n(n-1)$ parameters.

For example, for a $2\times 2$ ${\cal{H}}$, $V$ has one complex
mixing angle and one  complex phase (see (\ref{RV})).  For a
$3\times 3$ ${\cal{H}}$, $V$ has three complex mixing angles
and three  complex mixing phases.  It may be parameterized as
\begin{eqnarray}
V=
\left(\begin{array}{ccc}
1 & 0 & 0\\
0 & c_2 & s_2 e^{i\alpha_2}\\
0 & s_2 e^{-i\alpha_2} & -c_2\end{array}\right)
\left(\begin{array}{ccc}
c_1 & s_1 e^{i\alpha_1} & 0\\
-s_1 e^{-i\alpha_1} & c_1 & 0\\
0 & 0 & 1\end{array}\right)
\left(\begin{array}{ccc}1 & 0 & 0\\
0 & c_3 &s_3 e^{i\alpha_3}\\
0 & -s_3 e^{-i\alpha_3} & c_3\end{array}
\right),
\label{B9}
\end{eqnarray}
where $c_i=\cos\theta_i$ and $s_i=\sin\theta_i$ $(i=1,2,3)$.
$\theta_i$ and $\alpha_i$ are complex.



\begin{thebibliography}{[100]}
\bibitem{foot1}
             The eigenstates of propagation
             defined here are not physical as is
             evident from the fact that
             $\langle \phi'_a|\phi'_b\rangle \ne \delta_{ab}$.
             This relation follows because the $|\phi'_a\rangle$'s
             are related to the orthonormal states $|\phi_a\rangle$
             by a nonunitary transformation.
\bibitem{foot2}
             By construction, the orthonormal and complete conditions
             of $|\phi_a\rangle$ at $t=0$ are given by
             $\langle\phi_a|\phi_b\rangle=\delta_{ab}$ and
             $\sum_a|\phi_a\rangle\langle\phi_a|=1$. The situation
             for the propagation eigenstates is different \cite{Sachs1,EL,JL1}.
             For a given
             $|\phi'_a\rangle$, its
             conjugate state is
              $\langle\tilde{\phi}'_a|\ne \langle\phi'_a|$,
             where $\langle\phi_a'|=|\phi_a'\rangle^{\dag}$ as
             usual.
             The orthonormal and complete conditions at $t=0$ are
             $\langle\tilde{\phi}'_a|\phi'_b\rangle=\delta_{ab}$ and
             $\sum_a|\phi'_a\rangle\langle\tilde{\phi}'_a|=1$.
\bibitem{foot4}
             For simplicity we do not display finite
             vertex corrections explicitly.
\end{thebibliography}
\end{document}